# Valley-projected edge modes observed in underwater sonic crystals

Yuanyuan Shen[1], Chunyin Qiu[1*], Xiangxi Cai[1], Liping Ye[1], Jiuyang Lu[2], Manzhu Ke[1], and Zhengyou Liu[1,3]

[1]Key Laboratory of Artificial Micro- and Nano-structures of Ministry of Education and School of Physics and Technology, Wuhan University, Wuhan 430072, China

[2]School of Physics and Optoelectronic Technology, South China University of Technology, Guangzhou, Guangdong 510640, China

[3]Institute for Advanced Studies, Wuhan University, Wuhan 430072, China

**Abstract:** Recently, the topological physics in acoustics has been attracting much attention. However, all the studies are aimed to elastic or airborne sound systems. Realizing topological insulators for underwater sound is of great importance, since water is another crucial sound medium in addition to solid and air. Here we report an experimental study on the valley-projected edge states for underwater sound. The edge states are directly observed in our ultrasound scanning experiments, together with a solid evidence for the valley-selective excitation. The experimental data agree well with our numerical results. Prospective applications can be anticipated, such as for underwater sound signal processing and ocean noise control.

*Corresponding author.　cyqiu@whu.edu.cn

The discovery of the quantum Hall effect[1,2] opens the door to the field of peculiar topological phases of matter.[3-6] Recently, intense efforts have been devoted to realizing two-dimensional (2D) acoustic analogues[7-33] of topological insulators and the associated edge modes protected by symmetry. On the one hand, the macroscopic controllability enables the acoustic structures to be exceptional platforms to explore the intriguing topological properties that are challenging in original atomic systems. On the other hand, the topological edge states are particularly attractive to overcome some disorder-related restrictions in conventional acoustic technologies.

Different approaches have been proposed to design 2D acoustic topological insulators. The first approach mimics integer quantum Hall insulators with broken time-reversal symmetry. To break the time-reversal symmetry, gyroscopic structures[7,8] and circulating fluid flows[9-12] have been introduced to the mechanical and acoustic systems, respectively. However, practical experiments are challenging for such methods. Without breaking time-reversal symmetry, the second approach imitates quantum spin Hall insulators through constructing acoustic pseudospins. Despite the fact that different strategies have been used to construct pseudospins,[13-17] this approach often suffers complex structures or narrow operation bandwidth. Recently, the valley pseudospins proposed originally in solids,[18-22] which are degenerate freqency extremum states in momentum space,[23,24] have been considered to design acoustic topological insulators without breaking time-reversal symmetry.[25-33] Interestingly, the acoustic valley Hall (AVH) insulators are easy to fabricate and the operation bandwidth can be tailored by simply rotating the anisotropic scatterers.[25,27,28]

To the best of our knowledge, so far the investigations on acoustic topological insulators have been focusing on the artificial structures working for elastic waves[7,8,13,14,29-33] and airborne sound.[9-12,15-17,25-28] Water is another important sound carrier and the waterborne acoustics plays a momentous role in underwater communications, positioning, and telemetry. Therefore, it is of great interest to realize waterborne acoustic topological insulators. By exploiting the unique merits of the AVH insulators, here we report an experimental realization of the topological sound transport in water background. As illustrated in Fig. 1(a), the sonic crystal (SC) under consideration is made of a triangular lattice of identical meta-molecules. Each meta-molecule consists of three closely-packed steel cylinders. The AVH phase transition occurs at special orientations of the anisotropic meta-molecules, across which two topologically distinct AVH insulators appear. As a smoking gun evidence of the topological states of matter, the presence of nontrivial AVH edge modes has been validated experimentally. The valley-selective excitation of the edge modes has been confirmed by simple and complex interface systems. Our findings may provide a platform for designing exotic underwater sound devices.

Throughout this paper, all simulations are carried out by the commercial finite-element software (COMSOL MULTIPHYSICS). The material parameters used are: the mass density $\rho_s = 7760$ kg/m$^3$, the longitudinal velocity $v_l = 6010$ m/s, and the transverse velocity $v_t = 3230$ m/s for steel; and the mass density $\rho_0 = 1000$ kg/m$^3$ and the sound speed $c_0 = 1490$ m/s for water. Our ultrasonic experiments are performed in a big water tank. A pair of identical transducers (with diameter ~2.5 cm) is employed to measure transmission spectra.[34] To detect the field distribution of edge states, a needle-like transducer of diameter 1.2 mm is used for spatial scanning point-by-point.

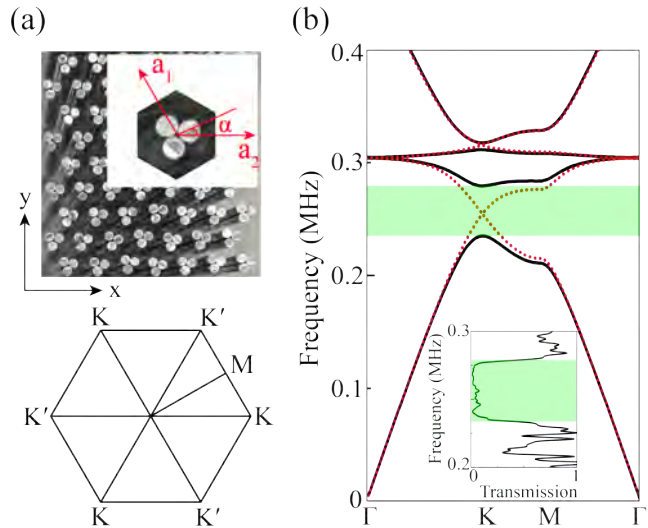

FIG. 1. (a) An image for the water-immersed SC (upper panel) made of a triangular lattice array of meta-molecules, together with its unit-cell geometry (inset) and associated first Brillouin zone (lower panel). Each meta-molecule is formed by three closely-packed steel cylinders, and the angle $\alpha$ characterizes its orientation. (b) Numerical band structure for the SC with $\alpha = 20^{\circ}$ (black solid line), comparing with the SC with $\alpha = 0^{\circ}$ (red dotted line). Inset: Power transmission spectrum measured experimentally for a 14-layer sample with surface normal along the ΓK direction, where the shadow region indicates the theoretical bandgap labeled in the band structure.

As shown in Fig. 1(a), our 2D SC consists of a triangular lattice array of meta-molecules immersed in water. Each meta-molecule is stacked closely by three identical steel cylinders. The lattice constant $a = 3.1$ mm and the cylinder diameter $d = 1.0$ mm. The rotation angle $\alpha$ characterizes the orientation of the anisotropic meta-molecule with respect to the +x axis. In our experiments, the steel cylinders are precisely arranged by threading them through a pair of plastic holey templates. Obviously, the meta-molecule has a spatial symmetry of $C_{3v}$. Similar to the previous works,[25,27,28] the SC property is closely related to the orientation of the meta-molecules. If $\alpha = n\pi/3$ (with $n$ being an arbitrary integer), the SC hosts conic degeneracy at the corners of the first Brillouin zone, $K$ and $K'$, due to the protection of $C_{3v}$ symmetry of the SC; otherwise ($\alpha \neq n\pi/3$), an omnidirectional bandgap opens because of the mismatch of the three mirrors between the lattice and the meta-molecule. These can be observed in Fig. 1(b), the numerical band structures for the SC with $\alpha = 0^{\circ}$ (red dashed line) and $\alpha = 20^{\circ}$ (black solid line). The sizable complete bandgap (from 0.235MHz to 0.279MHz) for $\alpha = 20^{\circ}$ has been confirmed by the experimental transmission spectrum (see inset). The band gap can be tuned even wider if the angle is increased to $\alpha = 30^{\circ}$.

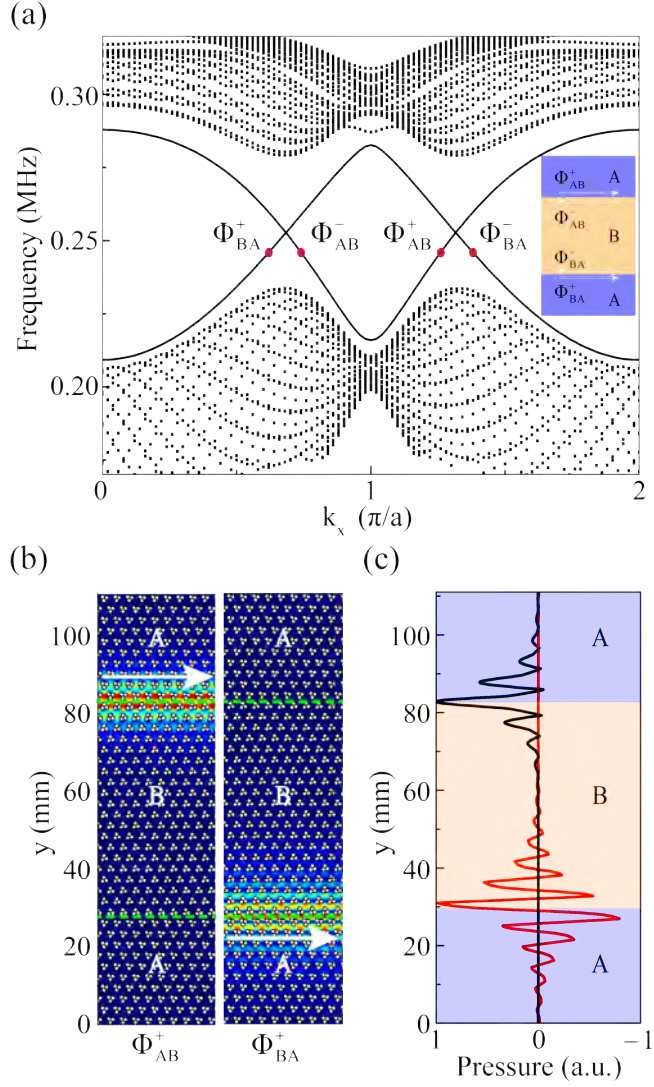

FIG. 2. (a) Dispersions for the SC interfaces along the $x$ direction, evaluated for an A-B-A super-lattice structure with fictitious periodicity applied in the $y$ direction. Inset: a schematic illustration for the super-lattice structure, where the white arrows indicate the traveling directions of the time-reversal pairs of the valley-projected edge states on each interface. (b) Eigen-field distributions (indicated by pressure amplitudes) for the two right-moving edge modes $\Phi_{AB}^+$ and $\Phi_{BA}^+$ (exemplified at 0.246 MHz). The SC interfaces are indicated by the horizontal green dashed lines. (c) Pressure distributions plotted along the vertical direction for the edge modes $\Phi_{AB}^+$ (black line) and $\Phi_{BA}^+$ (red line).

Interestingly, the SCs with opposite $\alpha$ values belong to topologically distinct AVH insulators, despite the fact that they carry completely identical band structures since these two SCs can be directly linked by a mirror operation with respect to the *x* axis. Hereafter we focused on the cases $\alpha = 20°$ and $\alpha = -20°$, referred to as SC-A and SC-B, respectively. According to the bulk-boundary correspondence,[25] each interface made of the SC-A and SC-B will host a time-reversal pair of the valley-projected edge states. To confirm this, we have simulated the interface dispersions by an A-B-A superlattice structure, associated with Bloch boundary condition used along the *x* direction and the superlattice periodicity applied along the *y* direction. As shown in Fig. 2(a), the simulation gives the edge modes for the A-B and B-A interfaces simultaneously, labeled by $\Phi_{AB}^{\pm}$ and $\Phi_{BA}^{\pm}$ respectively. Here the superscripts $\pm$ represent the group velocities of the edge modes along the $\pm x$ directions. The AVH edge modes are highly concentrated around the interfaces, as exemplified in Fig. 2(b) by two right-moving eigenstates $\Phi_{AB}^{+}$ and $\Phi_{BA}^{+}$. Interestingly, these horizontal edge states have strikingly different properties. As shown in Fig. 2(c), $\Phi_{AB}^{+}$ is locally symmetric with respect to the horizontal axis, whereas $\Phi_{BA}^{+}$ is locally antisymmetric with respect to the same axis. This difference is acceptable by intuition since the interfaces A-B and B-A have much different fine features inherently. As shown below, strikingly different sound responses to such two edge modes are observed experimentally.

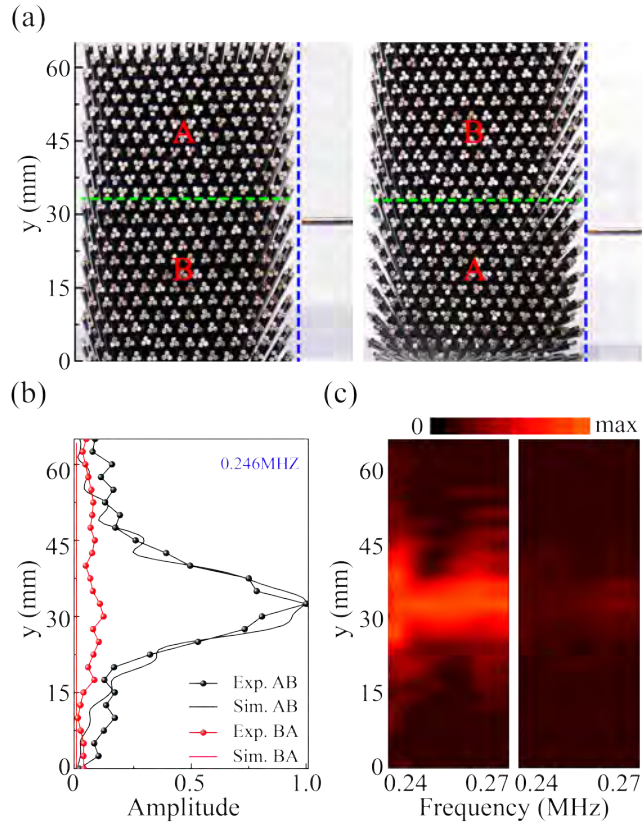

FIG. 3. (a) Photographs of the experimental samples with horizontal interfaces A-B (left) and B-A (right), respectively, both indicated by the green dashed lines. A Gaussian beam is launched toward the interface from the left hand side of the sample. (b) Transmitted pressure amplitude distributions measured (circles) and simulated (lines) along the vertical blue dashed lines indicated in (a). (c) Transmitted sound signals through the interfaces A-B (left) and B-A (right), scanned experimentally over the whole bulk gap region.

To confirm the valley-projected edge modes experimentally, we have designed two samples that form horizontal SC interfaces A-B and B-A separately [see Fig. 3(a)]. Both have a size of 65 cm x 40 cm, made of 336 meta-molecules in total. In each case, a Gaussian beam is launched horizontally from the left hand side of the sample, and the transmitted sound signal is scanned by a needle-like transducer along the vertical dashed line on the right hand side of the sample. Figure 3(b) exemplifies the transmitted pressure amplitude distributions at 0.246 MHz. Remarkably, the experimental data, in good agreement with the simulations, exhibit a weak sound signal transmitted through the interface B-A, in a sharp contrast to the well-excited interface A-B. As mentioned above, the strikingly different coupling efficiency originates from the locally symmetric and antisymmetric eigenstates $\Phi_{AB}^+$ and $\Phi_{BA}^+$ with respect to the incident beam. Interestingly, such remarkable contrast emerges in the entire bulk gap region [see Fig. 3(c)].

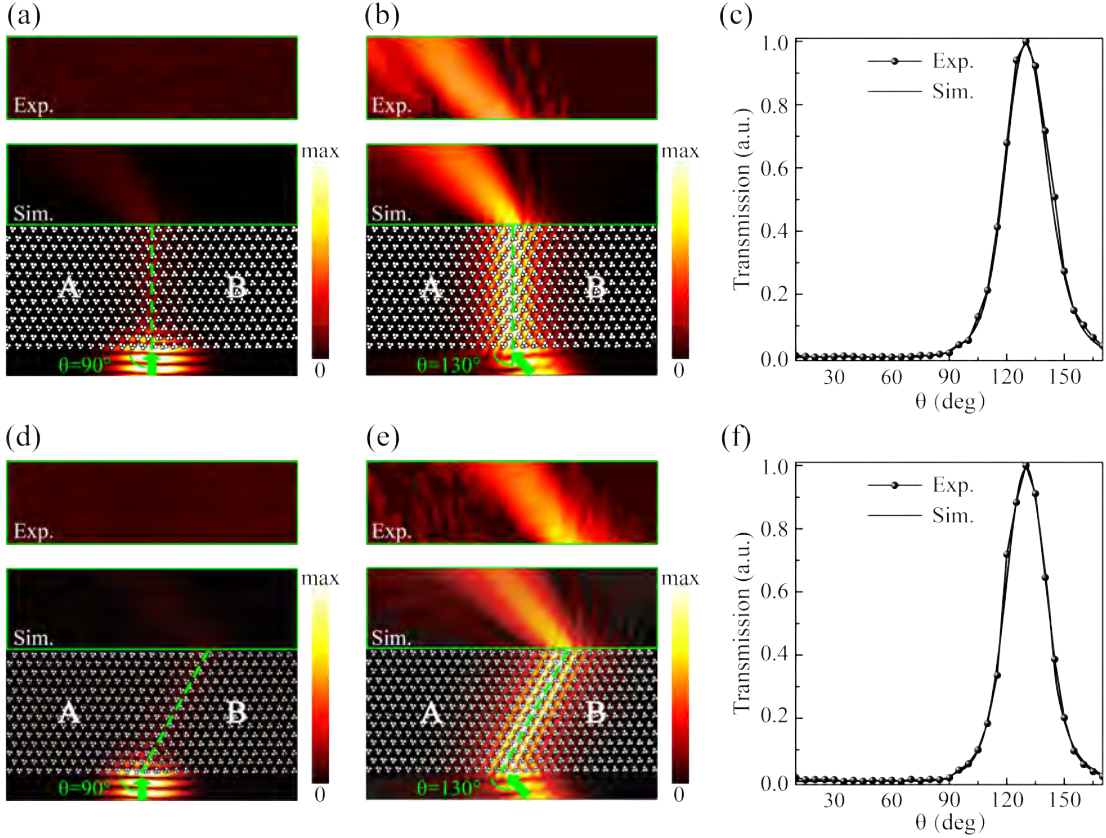

FIG. 4. Angularly selective excitation of the valley-projected edge states, where (a)-(c) correspond to a vertical interface, and (d)-(f) correspond to an interface inclined at $\pi/6$ with respect to the $y$ axis. (b) and (e) display the pressure amplitude distributions well excited by a Gaussian beam at the incident angle $\theta = 130^\circ$, in contrast to the cases (a) and (d) incident at $\theta = 90^\circ$. (c) and (f): $\theta$-dependent transmission spectra for the vertical and inclined interfaces. All simulations (lower panels), exemplified at 0.246 MHz, are well reproduced by our ultrasonic measurements (upper panels) at the output facets.

As pointed out in Ref. 25, the topological AVH edge state stems inherently from the single-valley physics. This leads to angularly selective excitation of the edge states according to the conservation of the momentum parallel to the sample boundary, i.e., $k_\| = k_0 \cos\theta$. Here $k_\|$ is the projection of K (or K′) point to the boundary of the sample, $k_0$ is the wave number in free space, and $\theta$ is the incident angle defined with respect to the +x direction. To confirm this property, we consider first a sample designed with a vertical interface. For comparison, in Figs. 4(a) and 4(b) we present the field patterns excited by the Gaussian beams with $\theta = 90°$ and $\theta = \cos(k_\|/k_0) \approx 130°$, respectively. It is observed that the AVH edge mode (projected by K valley here) is well excited at $\theta = 130°$ (associated with the output beam parallel to the incidence), in striking contrast to the deep suppression for the incidence at $\theta = 90°$. The experiments (upper panels) reproduce well the simulations (lower panels). The incidence sensitivity can be further confirmed by Fig. 4(c), the $\theta$-dependent power transmissions through the interface, where the optimized transmission occurs around the expected incident angle $\theta = 130°$. Interestingly, the phenomenon is irrelevant to the orientation of the interface. This can be seen in Figs. 4(d)-4(f) by the sample designed with an inclined interface, associated with almost invariant optimal incident angles. Note that both the cases in Fig. 3 satisfy the criterion of momentum conservation, while the interface state $\Phi_{BA}^+$ is forbidden due to the mismatched parity with respect to the incident wave.

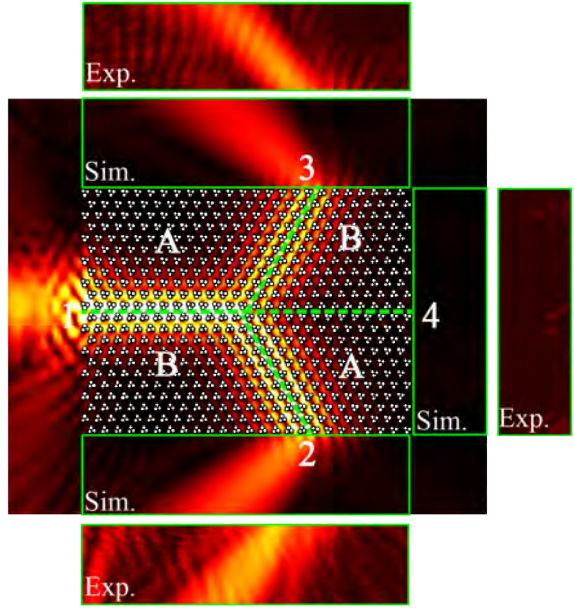

FIG. 5. Valley-selective sound transport in a four-port topological junction assembled with the SCs A and B alternately. The numerical and experimental field distributions are exemplified at 0.246 MHz.

The topological valley transport of sound may serve as the basis for designing devices with unconventional functions. For example, negligible backscattering has been observed in a sharply curved sound waveguide conceived by this mechanism.[25] Here we demonstrate an intriguing sound splitter built by different AVH insulators. As shown in Fig. 5, the device is constructed by the SC-A and SC-B alternately and has four ports in total. For a Gaussian beam incident normally upon the port 1, it is observed that the sound signal is well excited in the horizontal input waveguide (as expected in Fig. 3) and travels smoothly along the two side arms toward the ports 2 and 3, while is forbidden to transport along the horizontal output waveguide toward the port 4. The experimental data agree well with the numerical results. Again, such unusual sound splitting phenomenon is originated from the valley-projected nature of the AVH edge mode. As deduced from the theoretical model,[25] the forward-moving edge modes are always projected by the same valley $K$, as long as the SC-A locates at the left hand side of the SC-B as the propagation of sound. Specifically, the edge mode moving toward the port 4, projected by $K'$ valley otherwise, mismatches with the input edge mode and thus is strongly reflected, in contrast to the two well-excited side waveguides that host forward-moving edge modes projected by $K$ valley.

In summary, we have presented an experimental study of the valley-projected edge states for waterborne sound. The easy sample fabrication and the wide operation bandwidth are particularly useful in real applications. In additional to the angularly selective excitation of the edge modes, we have designed an intriguing sound splitter based on a four-port topological junction. Promising applications can be anticipated for such exceptional sound waveguides working in underwater environment.

**Acknowledgements**

This work is supported by the National Basic Research Program of China (Grant No. 2015CB755500); National Natural Science Foundation of China (Grant Nos. 11674250, 11774275, 11534013, 11747310); Natural Science Foundation of Hubei Province (Grant No. 2017CFA042).